# Understanding the phase stability in multi-principal-component AlCuFeMn alloy

Palash Swarnakar[1], M. Ghosh[2], B. Mahato[2], Partha Sarathi De[1], Amritendu Roy[1], *

[1]*School of Minerals, Metallurgical and Materials Engineering, Indian Institute of Technology Bhubaneswar, Jatni, Khurda-752050, Orissa, India*

[2] *Materials Engineering Division, CSIR-National Metallurgical Laboratory, Jamshedpur-831007, India*

**Abstract**
Method(s) that can reliably predict phase evolution across thermodynamic parameter space, especially in complex systems are of critical significance in academia as well as in the manufacturing industry. In the present work, phase stability in equimolar AlCuFeMn multi-principal-component alloy (MPCA) was predicted using complementary first-principles density functional theory (DFT) calculations, and *ab-initio* molecular dynamics (AIMD) simulations. Temperature evolution of completely disordered, partially ordered, and completely ordered phases was examined based on Gibbs free energy. Configurational, electronic, vibrational, and lattice mismatch entropies were considered to compute the Gibbs free energy of the competing phases. Additionally, elemental segregation was studied using ab-initio molecular dynamics (AIMD). The predicted results at 300K align well with room-temperature experimental observations using x-ray diffraction, scanning and transmission electron microscopy on a sample prepared using commercially available pure elements. The adopted method could help in predicting plausible phases in other MPCA systems with complex phase stability.

*Corresponding author: amritendu@iitbbs.ac.in

## 1. Introduction

Multi-Principal-Component Alloys (MPCA) with four or more principal elements (5-35 at. %), alternatively mentioned as high/medium entropy alloys, have attracted research attention in recent years owing to their promising properties [1,2]. Although, a significant literature pool exists on experimental investigations into phases, microstructure, and properties in these alloys, the unexplored compositional domain is enormous. For rapid assessment of composition space, novel approaches like combinatorial thin film deposition were proposed [3]. Nonetheless, a vital question on the design of novel MPCA persists, "for a given composition, can one foresee the phases present vis-à-vis crystallographic descriptions?" Such ability to predict the stabilities across the thermodynamic phase space in an alloy can permit one to design MPCAs with tailored properties opening up new vistas of engineering applications.

Traditionally, Density Functional Theory (DFT) calculations have long been in use to successfully predict the ground-state structure and properties of simple single-phase systems [4–6]. However, attempts to analyse phase stability in multi-phase, multi-component materials with site-disorders as in the multi-principal-component alloys (MPCAs) using DFT-based methods is few and far in between [7]. In fact, complementary experimental substantiation is essential since thermal and or mechanical history strongly influences the phase evolution. One such extensively investigated system is equimolar AlCuFeMn [8–10], and the present work uses the same as a test case to comprehend the phase stability intricacies using DFT-based methods. Additionally, annealed AlCuFeMn alloy exhibits impressive oxidation resistance [10] as well as significant aqueous corrosion resistance and mechanical strength (Unpublished data). Therefore, comprehensive analysis of phase stability as well as the structure property correlations in AlCuFeMn alloy is of technological significance.

Experimental investigations on as-cast AlCuFeMn by Dutta et al. [8] demonstrated three different BCC phases (BCC1: ~5.11 Å . BCC2: ~2.93 Å, and BCC3: ~5.93 Å) and one FCC phase with lattice parameter, ~5.85 Å. Annealing the as-cast alloy at 773 K for 80 h resulted in a Cu-rich BCC and Fe-rich FCC phases with lattice parameters of ~2.93 Å and ~5.86 Å [9]. Annealing at a higher temperature of 1173 K for 100 h resulted in Fe-rich BCC and Cu-rich FCC phases of lattice parameters, ~4.10 Å and ~3.65 Å besides a minor intermetallic phase [10]. Similar, Cu-rich $L2_1$ ($a \sim 5.95$ Å) and Fe-rich FCC ($a \sim 3.66$ Å) phases were reported in non-equimolar Al-Cu-Fe-Mn alloy [11–13]. Using the above published works as a reference, the present work performed first-principles DFT-based Gibbs free energy computation and *ab-initio* molecular dynamics study in equimolar AlCuFeMn alloy to predict the phase stability landscape vis-à-vis elemental segregation. The theoretical prediction was examined vis-à-vis room-temperature powder x-ray diffraction, and electron microscopy data on a sample prepared using commercially available pure elements.

## 2. Methodology

Phase stability is the manifestation of the competition among the probable materials states under given thermodynamic boundary conditions. Equilibrium phase evolution in bulk materials under given temperature-pressure conditions is determined by minimizing the Gibbs free energy. Thus, phase stability in an MPCA is governed by Gibbs free energy of mixing ($\Delta G_{mix}$) [14]:

$$\Delta G_{mix} = \Delta H_{mix} - T\Delta S_{mix} \qquad (1)$$

where $\Delta H_{mix}$ is the enthalpy of mixing, $\Delta S_{mix}$ entropy of mixing. $\Delta H_{mix}$ can be approximated as [15],

$$\Delta H_{mix} = \sum_{i=1, i\neq j}^{N} 4\Delta H_{mix}^{ij} c_i c_j \qquad (2)$$

where $\Delta H_{mix}^{ij}$ denotes the melting enthalpy of binary equiatomic $ij$ system, while $c_i$ and $c_j$ are concentrations of element $i$ and $j$, respectively. The $\Delta S_{mix}$ can have diverse origin [14]:

$$\Delta S_{mix} = \Delta S_{conf} + \Delta S_{lat} + \Delta S_{vib} + \Delta S_{el} + \Delta S_{mag} \qquad (3)$$

where $\Delta S_{conf}$ represents the configurational entropy, $\Delta S_{lat}$ is the entropic contribution due to the long-range periodic arrangement of atoms in crystalline solid or lattice mismatch entropy [16], $\Delta S_{vib}$, $\Delta S_{el}$ and $\Delta S_{mag}$ are vibrational, electronic and magnetic entropies, respectively. Further, in addition to the solid solutions (BCC, FCC, and HCP) [1,17,18], intermetallic compounds, laves phases [19], and or amorphous MPCAs [20] (bulk metallic glasses (BMGs)) are also reported. In this regard, contributions from entropy sources besides $\Delta S_{conf}$ are critical in stabilising phases in MPCA systems.

Based on the literature results for the AlCuFeMn system, phases considered in this work include disordered, partially ordered, and completely ordered solid solutions of different configurations within the cubic symmetry. Furthermore, possible elemental segregation within the prospective phases were investigated using *ab-initio* molecular dynamics (AIMD) simulation by studying the atomic interactions through the partial pair distribution function (PDF). The phase evolution predictions at 300 K were validated by room-temperature powder x-ray diffraction (pXRD) with Rietveld refinement, field emission scanning electron microscopy (FESEM) and transmission electron microscopy (TEM) on a vacuum induction melted sample prepared from elements with commercial purity. The next subsection, discusses the computations performed in details.

## 2.1 Computation details

The current study has been commenced with a review of previous literature to prepare a list of possible phases (see Supplementary Material: Table S1, Fig. S1). Based on these experimental details equimolar solid solutions of AlCuFeMn were constructed by randomly distributing equal number of the component atoms in the lattice sites of a 2 × 2 × 2 supercell (except in L2$_1$, where atoms were randomised within the 16-atom parent phase). The method of special quasi-random structures (SQS) [21], as implemented in the Alloy-Theoretic Automated Toolkit (ATAT) was used [22]. The above exercise allowed generation of 5 prospective completely disordered solid-solution phases of equimolar AlCuFeMn *viz.,* SQS BCC, SQS B2, SQS FCC, SQS L1$_2$, and SQS L2$_1$. Subsequent partial and/or complete ordering of the constituent atoms in above phases while maintaining equimolar composition allowed generation of additional phases (Table1). A summary of the prospective phases for first-principles calculations is provided in supplementary section (Section 3, Table S2 and S3) while schematic description is provided in Figure 1.

The Vienna *Ab-initio* Simulation Package (VASP) [23,24] was used to perform first-principles density functional theory (DFT) [25] computation of total energy of the phase in Table 1 using the generalized gradient approximation (GGA) [25]. The Perdew-Burke-Ernzerhof (PBE) exchange-correlation functional (GGA-PBE) was used [26] to treat the exchange and correlation contributions of the electrons to the Hamiltonian of the ion-electron system to solve the Kohn-Sham equation [27]. An optimised plane-wave cut-off energy of 550 eV was used. Projector augmented wave (PAW) [28] potentials comprising of eleven ($3d^{10}4s^1$) valence electrons for Cu, eight ($3d^74s^1$) for Fe, three ($3s^23p^1$) for Al, and seven ($3d^54s^2$) for Mn were used. Conjugate gradient algorithm [29] was used for structural relaxation. The Brillouin zone samplings were done using the Monkhorst-Pack [30] k-points with a grid density of $2\pi \times 0.03$ Å$^{-1}$. A self-consistent field (SCF) convergence criterion of $10^{-5}$ eV/atom and a maximum allowed force on the constituent atoms of 0.01 eV/Å were used. All the calculations were non-spin polarized and were performed at 0 K.

Total energy computation using DFT allows estimation of the formation energy of the relaxed phases using:

$$E_{form} = \left( E_{AlCuFeMn}^{sys} - \left( \sum_i c_i E_i^{bulk} \right) \right) \quad (4)$$

where $E_{AlCuFeMn}^{sys}$ corresponds to the total energy of the AlCuFeMn system (eV/atom) at 0 K, $E_i^{bulk}$ is the total energy of the specie $i$ (eV/atom) (where $i$ = Al, Cu, Fe, Mn) in their bulk crystalline states at 0 K and $c_i$ is the concentration of species $i$. To establish the robustness of calculation, formation energy computation was repeated on larger supercells, *viz.*, $4 \times 4 \times 4$ supercells of SQS BCC, SQS B2, B2-3, $3 \times 3 \times 3$ supercells of SQS FCC, SQS L1$_2$, L1$_2$-1, $2 \times 2 \times 2$ supercells of SQS L2$_1$, and QH comprising more than 100 atoms (see Supplementary Material: Table S4). A similar trend in the formation energy for both small and large cells was noted and therefore all further calculations were continued on $1 \times 1 \times 1$ cell to save computation time.

The phase with the lowest formation energy is presumed to evolve, provided it is dynamically stable as characterized by the absence of any "soft mode"in the phonon band structure [31] computed using density functional perturbation theory (DFPT) [32] as implemented in PHONOPY [31]. The relative stability of the competing phases at finite temperatures was determined by comparing the Gibbs free energy as a function of temperature. Under the harmonic approximation, the Helmholtz free-energy $F(V,T)$, as a function of volume and temperature, can be expressed as:

$$F(V,T) = E^{sys}_{AlCuFeMn}(V) + F_{vib}(V,T) + F_{el}(V,T) - T(S_{conf} + S_{lat} + S_{mag}) \quad (5)$$

where $E^{sys}_{AlCuFeMn}$ is the total free energy at 0 K, $F_{el}(V,T)$ is the electronic free energy, $F_{vib}(V,T)$ is the vibrational free energy, $S_{conf}$, $S_{lat}$ and $S_{mag}$ are entropies concerning to configurational, lattice mismatch and magnetism of a system, respectively. Gibbs free energy was determined from the Helmholtz free energy [33],

$$G(p,T) = \min_{V}[F(V,T) + pV] \quad (6)$$

For completely disordered phases, the configurational entropy term is $\Delta S_{conf} = -k_B \sum_i c_i \ln c_i$, where $k_B$ is the Boltzmann constant, and $c_i$ is the concentration of element $i$. $\Delta S_{conf}$ is expected to be modified to a certain degree for partially ordered phases. Based on the sub-lattice model [34], the configurational entropy of partially ordered phases could be calculated as:

$$\Delta S^{PO}_{conf} = -R \left[ \frac{\sum_{x=1}^{x} a^x \sum_{i=1}^{N} f_i^x (\ln(f_i^x))}{\sum_{x=1}^{x} a^x} \right] \quad (7)$$

where PO stands for partially ordered phase, $R$ is the gas constant, $a^x$ and $f_i^x$ denote the number of sites and the fraction of element of species $i$ on the sublattice $x$, and $N$ is the total number of species $i$. The vibrational free energy of the system was computed using the canonical distribution within the harmonic approximation [31,32]. However, it should be noted that crystal potential is an anharmonic volume function. Computation of anharmonic contribution to the total phonon energy is computationally expensive. To account for the anharmonic effect to the total phonon energy in an average manner, quasi-harmonic approximations were applied where the energy-volume curves ($E - V$) along with the phonon energies were fitted with the third-order Birch-Murnaghan equation of states (EOS) [35] on seven selected volumes (0.95, 0.97, 0.99, 1.00, 1.01, 1.03, and 1.05 of equilibrium volume ($V_0$)). Furthermore, the zero-point energy (ZPE) at 0 K, was included in $F_{vib}$ thus excluding separate computations. The electronic free energy ($F_{el}$) was computed from electronic eigen energies [36] using the PHONOPY [31,33] during QHA calculations. The lattice mismatch entropy of mixing $S_{lat}$ (from Eq. 3) is a function of concentration $c_i$, atomic diameter $d_i$, and packing efficiency $\xi$, and was computed using the equations provided by Ye et al. [16].

Elemental segregations in AlCuFeMn was studied using *ab-initio* molecular dynamics (AIMD) simulations as implemented in VASP where Nosé algorithm [37,38] was used to equilibrate the structure in the canonical ensemble (N, V, T). A cubic volume (SQS SC) (volume = 2541.12 Å$^3$) with randomly distributed 100 atoms in equiatomic proportion (i.e., 25 atoms of each element) was used to eliminate any presumption on the parent phase. The SQS SC phase was annealed at 2273 K for 15 ps during simulation with a time interval of 1 fs. The partial PDF ($g_{\alpha\beta}$) calculated using the following equation [14]:

$$g_{\alpha\beta}(r) = \frac{V}{N_\alpha N_\beta} \frac{1}{4\pi r^2} \sum_{i=1}^{N_\alpha} \sum_{j=1}^{N_\beta} \langle \delta(|r_{ij} - r|) \rangle \qquad (8)$$

where $V$ is the volume of the supercell, $N_\alpha$ and $N_\beta$ are the numbers of elements $\alpha$ and $\beta$, $|r_{ij}|$ is the distance between elements $\alpha$ and $\beta$, and <> is the time average of different configurations.

**2.2. Experimental Details**

Bulk AlCuFeMn alloy (from elements/ferroalloys of commercial purity) was melted and cast using vacuum induction melting and casting system at a controlled pressure of 10$^{-4}$ mbar. The cast block was annealed at 900°C for 100 hours. The microstructure and chemical composition were investigated using field emission-scanning electron microscopy (FESEM) and energy dispersive spectroscopy (EDS). Metallographic polishing of all samples was accomplished using SiC papers and diamond paste, while the finishing was performed using 0.05 $\mu$m colloidal silica. The FESEM studies were performed using a back-scattered electron (BSE) detector with an acceleration voltage of 20 kV. Room-temperature powder x-ray diffraction (pXRD) pattern of the alloy was acquired with Cu k$\alpha$ ($\lambda$ = 1.5418 Å) radiation using the Bragg-Brentano geometry. The FullProf Suite software [39] was employed for Rietveld refinement of the room-temperature pXRD data using the structural models predicted by the DFT study. The presence of ordered phases was verified using by conventional analytical transmission electron microscopy (TEM).

## 3. Results

*3.1 Temperature evolution of the prospective phases*

For all the probable phases listed in the Supplementary Material: Table S2 and S3, the formation energy computed using Eq. 4, is presented in Table 1. Assuming that the phases with negative formation energy will prevail over those with positive formation energy, subsequent studies were performed only on those phases having negative formation energy. The computed phonon dispersion in the first Brillouin zone for all the phases with negative formation energy are shown in Figure 2. Due to the complexity of the crystal structures, the same crystal dimension for phonon calculations has been used. Nonetheless, to check the consistency, we also calculated the phonon spectra for the supercell dimension of $2 \times 2 \times 2$ in the QH phase (not shown for brevity). It has been observed that $L1_2$-4, $L2_1$-2, and $L2_1$-4 phases possess negative phonon frequencies, implying structural instability (see Figure 2). As the number of atoms per lattice site increases, the structural symmetry decreases. Due to lower structural symmetry at the lattice sites, the number of phonon bands (which is proportional to the number of atoms in the cell) is higher in disordered SQS phases such as BCC, B2, FCC, $L1_2$, and $L2_1$. On the contrary, owing to their higher structural symmetry, partially ordered B2 type phases, B2-2 and B2-3 phases, exhibit fewer phonon bands (see Fig. 2 (g, h)). Similarly, the ordered QH phase, which has the highest symmetry among the studied phases, has the fewest phonon bands (see Fig. 2 (f)). In Table 2 the configurational (Eq. 7) and lattice mismatch (see supplementary material: Eq. S4) entropies for only the dynamically stable phases are listed. In Figure 3 the Gibbs free energy of the dynamically stable phases (given in Table 2) with respect to temperature is plotted to assess the relative stability and probable phase transitions.

*3.2. AIMD simulation*

The approximated first nearest-neighbour (FNN) peaks were determined from Figure 4 and are tabulated in Table 3. Two distinct peaks are observed in Figure 4, where Fe-Fe, Fe-Mn, Mn-Mn, Cu-Cu, Cu-Fe, and Cu-Mn share a mean FNN of 2.27 Å (FNN-1) and Al-Cu, Al-Fe, Al-Mn, Cu-Cu, Cu-Fe, and Cu-Mn share a mean FNN of 2.52 Å (FNN-2). The peak of FNN-1 is dominated by Mn-Mn interactions, followed by Fe-Fe and Fe-Mn interactions.

*3.3. Structural and microstructural Characterisation of cast and annealed AlCuFeMn alloy*

Room-temperature pXRD pattern of the bulk alloy was refined using the Rietveld method [40] as implemented in the FullProf suite [39]. The DFT predicted phases at 300 K, *viz.*, B2-3, $L2_1$-3, and $L1_2$-1 were used to refine the pXRD data. The pXRD and the corresponding refinement are shown in Figure 5. The refined lattice parameters and other crystallographic details are listed in Table 5.

The microstructure of the alloy was studied using a field emission scanning electron microscope (FESEM) with a back-scattered electron (BSE) detector, and the average chemical composition was determined using an energy dispersive spectroscopy (EDS) attachment with the FESEM. The back-scattered image in Figure 6 (a), and the corresponding EDS elemental mapping, Figure 6 (b), figure out two regions *viz.,* copper-rich bright strips, and iron-manganese enriched dark matrix. Aluminium is uniformly distributed throughout the alloy (Figure 6 (b)). To understand the long-range atomic arrangement in the copper-rich and Fe-Mn rich regions, transmission electron microscopy was carried out on same specimen. With reference to FESEM micrograph in Figure 6, two distinct regions have been also revealed in TEM micrograph (Figure 7 (a)), as NBD1 and NBD2. Point EDS analysis of these two regions shown in Fig. 7(c) and 7(e), exhibited, that the region labelled as NBD1 is copper-rich while that of labelled as NBD2 is iron-manganese-rich. The selected area diffraction pattern (SADP) in Fig. 7 (b, d), affirmed that Cu-rich NBD1 region has ordered BCC structure *i.e.*, B2 structure while Fe-Mn rich NBD2 region possesses ordered FCC structure i.e., $L2_1$. High-magnification dark-field TEM micrographs in Fig. 7 (g)-(h) of NBD2 region demonstrates nano-sized precipitates. Nano-diffraction of these precipitates, shown in Fig. 7 (f), confirmed $L1_2$ structure of precipitates.

## 4. Discussions

From the first-principles study, it was noticed that the differences in the Gibbs free energy among the competing phases are not dramatic, suggesting that more than one phase might coexist in the AlCuFeMn MPCA. It is to be noted that the anharmonic contribution in Gibb's free energy calculation was ignored in the current work to reduce the computation time and instead "quasi-harmonic approximation" was used. While such an assumption would inevitably incorporate some error in the computation of Gibb's free energy, the subsequent experimental study shows that the said error is within the allowable limit and did not alter the phase stability. Utilization of quasi harmonic approximation to compensate for the anharmonic contribution was successfully employed to predict phase stability in open domain literatures [41–43].

At temperatures ≤200 K the phases with increasing energies are B2-3 < $L2_1$-3 < B2-2 < QH < $L1_2$-1 , as shown in Figure 3 (a). The maximum energy difference between the four lowest-energy structures (B2-3, $L2_1$-3, B2-2, QH, and $L1_2$-1) is 0.017 eV/atom, which is within the range of thermal fluctuation at 200 K, thus suggesting that these phases may coexist. At 180 K, the energy of the QH phase becomes higher than that of $L1_2$-1. At temperatures ~ 300 K, the phases with increasing energies are B2-3 < B2-2 < $L2_1$-3 < $L1_2$-1, as shown in Figure 3 (a) with energy difference (with respect to the lowest-energy phase at that temperature) within the range of thermal fluctuation at 300 K, thus suggesting the co-existence of these phases at room-temperature. Iso-structural B2-2 and B2-3 are distinguishable only at the sub-lattice level with B2-3 having lower free energy than B2-2; therefore, B2-2 was ignored during the experimental study. Disordered phases, *viz.*, SQS BCC/B2, and SQS FCC/L1$_2$, are stable at higher temperatures (~1000 K), Figure 3 (b). Therefore, only three phases are predicted at room-temperature, *viz.*, B2-3, L2$_1$-3, and L1$_2$-1.

During the room-temperature pXRD measurement, the sample was manually rotated and adjusted until all the peaks were detected. From the first-principles study in the preceding sections, it was predicted that at 300 K, equimolar AlCuFeMn MPCA would have three phases, *viz.*, B2-3, L2$_1$-3, and L1$_2$-1 wherein B2-3 possesses $Pm\bar{3}m$ symmetry, L2$_1$-3 has $Fm\bar{3}m$ symmetry, and L1$_2$-1 crystallizes in $Pm\bar{3}m$ symmetry. The pXRD peaks at ~ 26º and 30º are associated with the 111 and 200 reflections from the L2$_1$-3 phase. The 30º peak also coincides with the super-lattice reflection from plane 100 of the B2-3 phase. The 111 reflection only differentiates the L2$_1$-3 phase from B$_2$-3, while the rest of the peaks overlap. Although the first-principles-based predictions suggested a partially ordered L1$_2$ phase, the pXRD could not capture the super-lattice reflections presumably due to very low intensity of the superlattice line. However, the SADP from the Cu-rich region reveals low-intensity superlattice spots indicating ordering in the BCC phase rendering it to be represented as, B2 phase. Superlattice spots in the Fe-Mn-rich region suggest ordering in the FCC phase; accordingly, it is labelled as L2$_1$. Similar observation of ordering in MPCAs was also noted in earlier studies [44–46].

The AIMD simulations also provided insight into elemental segregation through PDF analysis. SQS SC structure, annealed at 2273 K for 15 ps during simulation, reveals two phases, one of which is Fe-Mn-rich and the other is Cu-rich. From the PDF analysis, it was found, that due to the higher peak intensity of Mn-Mn, and Fe-Mn in FNN-1, a Fe-Mn-enriched phase with a preference for Mn segregation could be expected [47]. In the FNN-2 peak, Al-Fe interactions have the highest peak intensity, followed by Al-Cu, Cu-Cu, Al-Mn, Cu-Fe, and Cu-Mn. Despite Al-Fe having the highest intensity in FNN-2, Cu interactions with self and others are dominant; therefore, FNN-2 indicates a different phase, particularly rich in Cu. Al-Al, with an FNN of 2.81 Å, does not share a peak with any other element, and the Al-Al interactions are weak (compared to FNN-1 and FNN-2); therefore, Al prefers to bond with other elements rather than itself [47]. In other words, Al should be present uniformly in the matrix.

In fact, FESEM images in back scattered mode revealed, that the microstructure consists of two major regions, Cu-rich region, and Fe-Mn-rich region (see Fig. 5). The distribution of Al within the matrix is consistent with the theoretical predictions. The phase-wise elemental distribution was taken from a BSE image of ∼ 12 $\mu$m × 12 $\mu$m, whereas the overall distribution of elements was taken from a BSE image of ∼ 300 $\mu$m × 300 $\mu$m. Cu-rich region comprises $of$ ∼ 47 at % Cu, followed by ∼ 23 at % Al, with each Fe and Mn containing ∼ 15 at %. Fe-Mn-rich region contains nearly 36 at % Fe, followed by equimolar Al and Mn (each having ∼ 25 at %) and Cu ∼ 14 at %. Nano-sized second phase L1$_2$ with low volume fraction was found to precipitate on the parent matrix of Fe-Mn-rich region. In Table 5, the elemental composition, determined by TEM (-EDS) was found to vary from that determined by FESEM (-EDS) shown in Table 4. This could be attributed to small area (point EDS) of a thin sample chosen in case of TEM-EDS study in comparison to FESEM-EDS that acquired x-ray signals from a large volume of the bulk sample.

**Conclusion**

The present study aims to rationalise the evolution of different phases in the multi-principal-component AlCuFeMn alloy using complementary first-principles density functional theory calculations as well as *ab-initio* molecular dynamics simulations. In this study, completely disordered, partially ordered, and completely ordered phases in equimolar AlCuFeMn MPCA were examined. In the DFT studies, configurational, electronic, vibrational, and lattice mismatch entropies were considered to compute the Gibbs free energies of the competing phases. Magnetic contributions were not considered in present investigation. Room-temperature pXRD, together with Rietveld refinement, TEM and SADP, bolstered the computational prediction by identifying three different crystallographic phases. Thus, theoretical and experimental analyses have led to the identification of two major phases at room-temperature (300 K), *viz.*, Cu-rich B2 (Al, Mn)$_{0.5}$(Cu, Fe)$_{0.5}$, and Fe-Mn-rich L2$_1$ (Al, Mn)$_2$(Cu)$_1$(Fe)$_1$, and a nano-sized low volume fraction L1$_2$ (Al)(Cu, Fe, Mn)$_3$ phase within Fe-Mn-rich matrix. It is believed that similar approach may help in predicting complex phase evolution in other multi-principal-component alloy systems.

**Supplementary Material**

Please refer to the supplementary material for further details.

**Acknowledgements**

PS thanks the Ministry of Education, Government of India (GOI) for fellowship. This work was supported by SERB, Govt. of India through project no. CRG/2019/00184.

**Data Availability**

The data that support the findings of this study are available from the corresponding author upon reasonable request.

# Tables and Figures

Table 1: Compilation of probable phases with their computed total energy $E^{sys}_{AlCuFeMn}$ (eV/atom) and formation energy $E_{form}$ (eV/atom) in AlCuFeMn MPCA.

| Orderings | System | Notation | $E^{sys}_{AlCuFeMn}$ (eV/atom) | $E_{form}$ (eV/atom) | Dynamically Stable? |
|---|---|---|---|---|---|
| Com Dis (SQS) | SQS BCC | — | -6.1278 | -0.0698 | Yes |
| | SQS B2 | — | -6.1278 | -0.0698 | Yes |
| | SQS FCC | — | -6.1142 | -0.0563 | Yes |
| | SQS L1$_2$ | — | -6.1187 | -0.0608 | Yes |
| | SQS L2$_1$ | — | -6.1032 | -0.0452 | Yes |
| Par Ord | B2 (Al,Cu)$_{0.5}$(Fe,Mn)$_{0.5}$ | B2-1 | -6.0145 | 0.0434 | — |
| | B2 (Al,Fe)$_{0.5}$(Cu,Mn)$_{0.5}$ | B2-2 | -6.1587 | -0.1008 | Yes |
| | B2 (Al,Mn)$_{0.5}$(Cu,Fe)$_{0.5}$ | B2-3 | -6.1734 | -0.1155 | Yes |
| | L1$_2$ (Al)$_1$(Cu,Fe,Mn)$_3$ | L1$_2$-1 | -6.1446 | -0.0867 | Yes |
| | L1$_2$ (Cu)$_1$(Al,Fe,Mn)$_3$ | L1$_2$-2 | -6.1135 | -0.0556 | Yes |
| | L1$_2$ (Fe)$_1$(Al,Cu,Mn)$_3$ | L1$_2$-3 | -6.0755 | -0.0176 | Yes |
| | L1$_2$ (Mn)$_1$(Al,Cu,Fe)$_3$ | L1$_2$-4 | -6.0704 | -0.0125 | No |
| | L2$_1$ (Al,Cu)$_2$(Fe)$_1$(Mn)$_1$ | L2$_1$-1 | -5.9933 | 0.0646 | — |
| | L2$_1$ (Al,Fe)$_2$(Cu)$_1$(Mn)$_1$ | L2$_1$-2 | -6.1167 | -0.0588 | No |
| | L2$_1$ (Al,Mn)$_2$(Cu)$_1$(Fe)$_1$ | L2$_1$-3 | -6.1299 | -0.0720 | Yes |
| | L2$_1$ (Cu,Fe)$_2$(Al)$_1$(Mn)$_1$ | L2$_1$-4 | -6.1355 | -0.0776 | Yes |
| | L2$_1$ (Cu,Mn)$_2$(Al)$_1$(Fe)$_1$ | L2$_1$-5 | -6.1316 | -0.0737 | No |
| | L2$_1$ (Fe,Mn)$_2$(Al)$_1$(Cu)$_1$ | L2$_1$-6 | -5.9894 | 0.0686 | — |
| Com Ord | Quaternary Heusler | QH | -6.1541 | -0.0962 | Yes |

Com Dis: Completely Disordered
Par Ord: Partially Ordered
Com Ord: Completely Ordered

Table 2: Computed formation energies and configurational and lattice mismatch entropies for dynamically stable phases in AlCuFeMn MPCA.

| System | $E^{sys}_{AlCuFeMn}$ (eV/atom) | $E_{form}$ (eV/atom) | $S_{conf}$ ($k_B$/atom) | $S_{lat}$ ($k_B$/atom) |
|---|---|---|---|---|
| SQS BCC | -6.1278 | -0.0698 | 1.386 | 0.082 |
| SQS B2 | -6.1278 | -0.0698 | 1.386 | 0.082 |
| SQS FCC | -6.1142 | -0.0563 | 1.386 | 0.131 |
| SQS L1$_2$ | -6.1187 | -0.0608 | 1.386 | 0.131 |
| SQS L2$_1$ | -6.1032 | -0.0452 | 1.386 | 0.131 |
| B2-2 | -6.1587 | -0.1008 | 0.693 | 0.082 |
| B2-3 | -6.1734 | -0.1155 | 0.693 | 0.082 |
| L1$_2$-1 | -6.1446 | -0.0867 | 0.824 | 0.131 |
| L1$_2$-2 | -6.1135 | -0.0556 | 0.824 | 0.131 |
| L1$_2$-3 | -6.0755 | -0.0176 | 0.824 | 0.131 |
| L2$_1$-3 | -6.1299 | -0.0720 | 0.347 | 0.131 |
| L2$_1$-4 | -6.1355 | -0.0776 | 0.347 | 0.131 |
| QH | -6.1541 | -0.0962 | 0.000 | 0.131 |

Table 3: Simulated FNN of the SQS SC phase obtained using AIMD at 2273 K for 15 ps in the AlCuFeMn MPCA.

| Pairs | FNN (Å) | Mean FNN (Å) |
|---|---|---|
| Al-Al | 2.81 | 2.81 |
| Al-Cu | 2.49 | |
| Al-Fe | 2.52 | |
| Al-Mn | 2.55 | 2.52 |
| Cu-Cu | 2.52 | |
| Cu-Fe | 2.53 | |
| Cu-Mn | 2.49 | |
| Fe-Fe | 2.24 | |
| Fe-Mn | 2.30 | 2.27 |
| Mn-Mn | 2.27 | |

Table 4: Elemental distribution obtained from EDS analysis on the experimentally observable regions in the AlCuFeMn MPCA.

| Region | Al (at. %) | Cu (at %) | Fe (at %) | Mn (at %) | Volume Fraction (%) |
|---|---|---|---|---|---|
| Cu-rich | 22.62 ± 0.67 | 47.02 ± 5.13 | 14.59 ± 2.96 | 15.76 ± 1.75 | 30 ± 1.5 |
| Fe-Mn-rich | 25.00 ± 0.49 | 12.90 ± 2.54 | 36.06 ± 2.12 | 26.00 ± 0.86 | 70 ± 2.3 |
| Overall | 23.42 ± 0.18 | 26.03 ± 0.17 | 27.72 ± 0.06 | 22.84 ± 0.21 | |

Table 5: Phases observed in the room temperature pXRD, with experimental (DFT computed) lattice parameters, refined site occupancy, and estimated phase composition in AlCuFeMn MPCA.

| Phase | Space Group | Exp. $a=b=c$ (DFT) (Å) | Phase fraction (wt %) | Atomic coordinates | Occupancy |
|---|---|---|---|---|---|
| B2-3 | $Pm\bar{3}m$ (221) | 2.91 (2.82) | 26.31 | Al 1a (0, 0, 0) | 0.50 |
| | | | | Mn 1a (0, 0, 0) | 0.50 |
| | | | | Cu 1b ($\frac{1}{2}, \frac{1}{2}, \frac{1}{2}$) | 0.70 |
| | | | | Fe 1b ($\frac{1}{2}, \frac{1}{2}, \frac{1}{2}$) | 0.30 |
| L1$_2$-1 | $Pm\bar{3}m$ (221) | 3.68 (3.62) | 7.41 | Al 1a (0, 0, 0) | 1.00 |
| | | | | Cu 3c (0, $\frac{1}{2}, \frac{1}{2}$) | 0.20 |
| | | | | Fe 3c (0, $\frac{1}{2}, \frac{1}{2}$) | 0.40 |
| | | | | Mn 3c (0, $\frac{1}{2}, \frac{1}{2}$) | 0.40 |
| L2$_1$-3 | $Fm\bar{3}m$ (225) | 5.87 (5.75) | 66.28 | Cu 4a (0, 0, 0) | 1.00 |
| | | | | Fe 4b ($\frac{1}{2}, \frac{1}{2}, \frac{1}{2}$) | 1.00 |
| | | | | Al/Mn 8c ($\frac{1}{4}, \frac{1}{4}, \frac{1}{4}$) | 0.50/0.50 |
| | | | | Al/Mn 8c ($\frac{3}{4}, \frac{3}{4}, \frac{3}{4}$) | 0.50/0.50 |

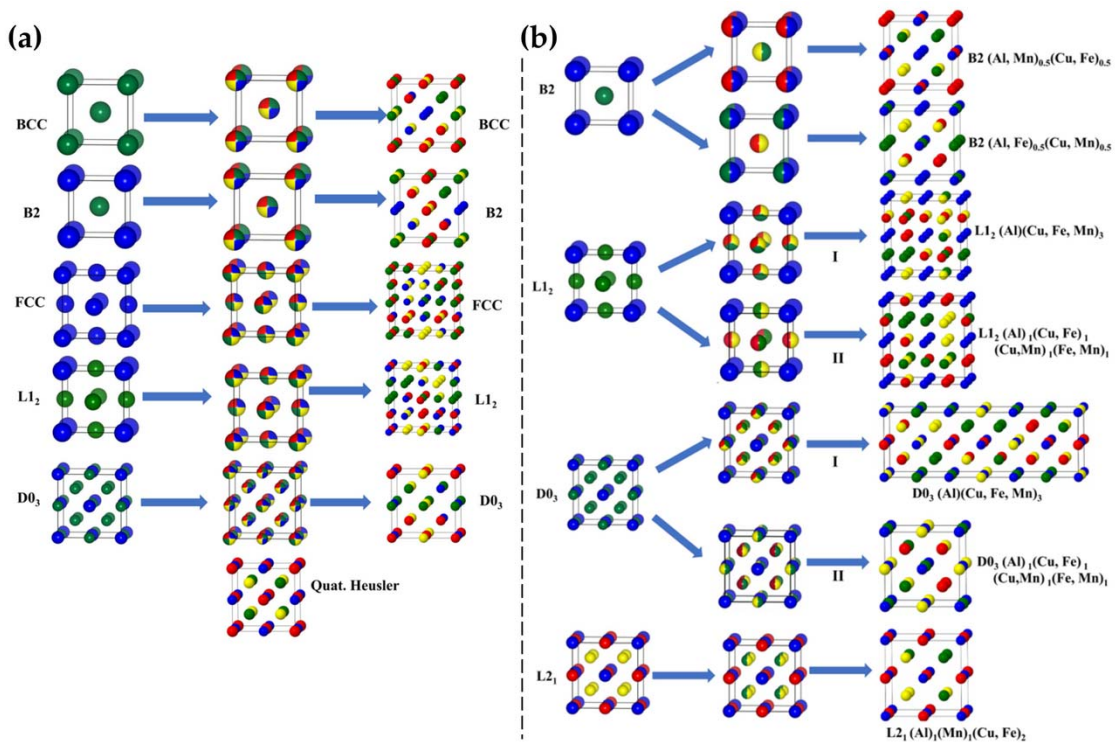

*Figure 1:* (a) Schematic representations of the parent phases generated through the method of SQS, and (b) configurations generated through partial lattice site occupancies in AlCuFeMn MPCA. Al, Cu, Fe and Mn atoms are represented by blue, yellow, green and red solid balls, respectively.

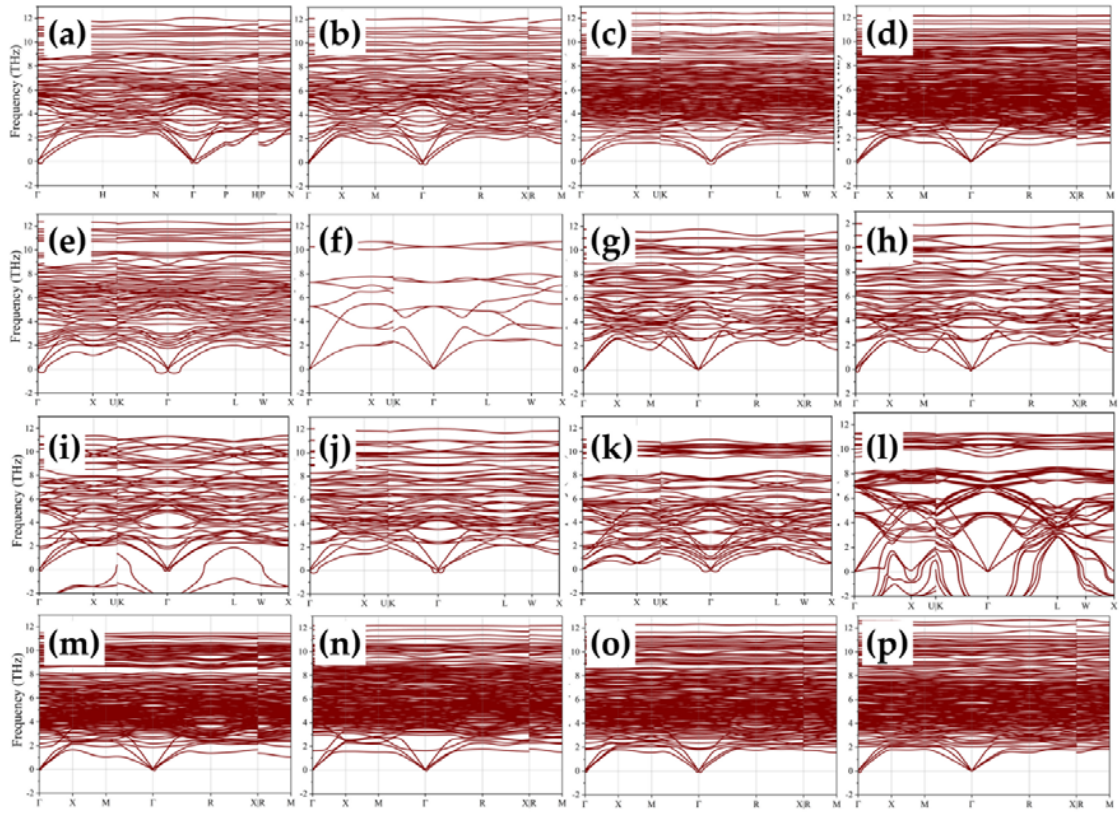

Figure 2 : The phonon bandstructure of (a) SQS BCC, (b) SQS B2, (c) SQS FCC, (d) SQS L12, (e) SQS L21, (f) QH, (g) B2-2, (h) B2-3, (i) L2$_1$-2, (j) L2$_1$-3, (k) L2$_1$-4, (l) L2$_1$-5, (m) L1$_2$-1, (n) L1$_2$-2, (o) L1$_2$-3, and (p) L1$_2$-4 structures in AlCuFeMn MPCA.

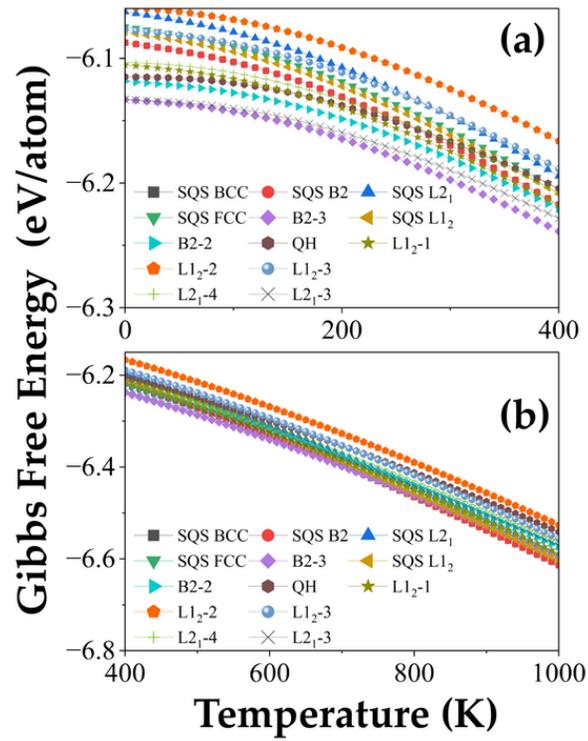

Figure 3: Comparative study of the Gibbs free energies among the dynamically stable phases at the temperature intervals of (a) 0-400 K, (b) 400-1000 K in AlCuFeMn MPCA.

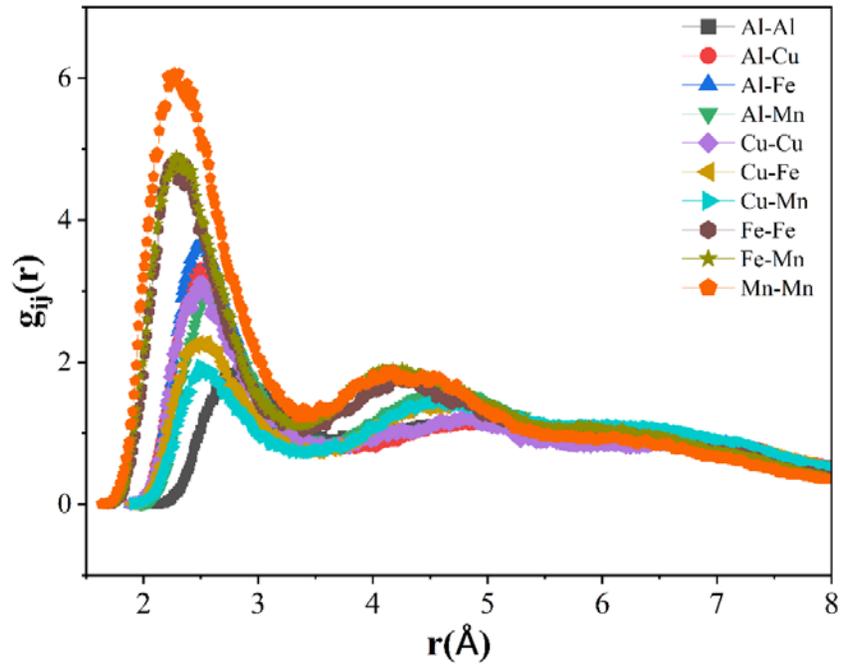

Figure 4: Partial pair distribution function (PDF) of SQS SC for 15 ps at 2273 K in AlCuFeMn MPCA.

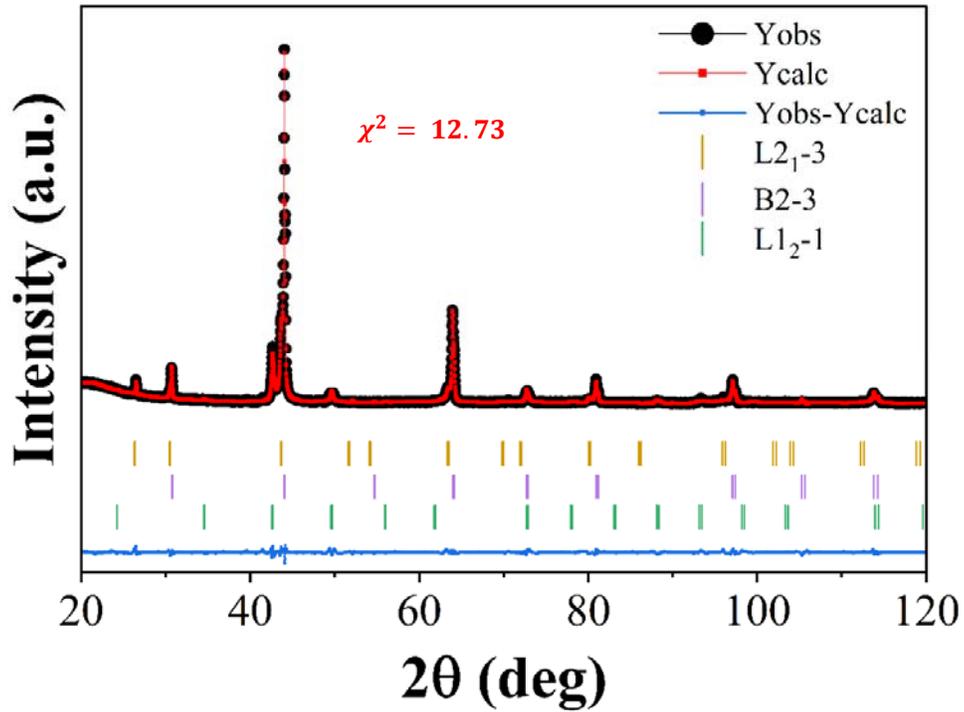

Figure 5: Powder x-ray diffraction and the corresponding Rietveld refinement in the AlCuFeMn MPCA. The goodness of the fitting ($\chi^2$) was ~12.73 with $R_{exp}$ ~ 4.41.

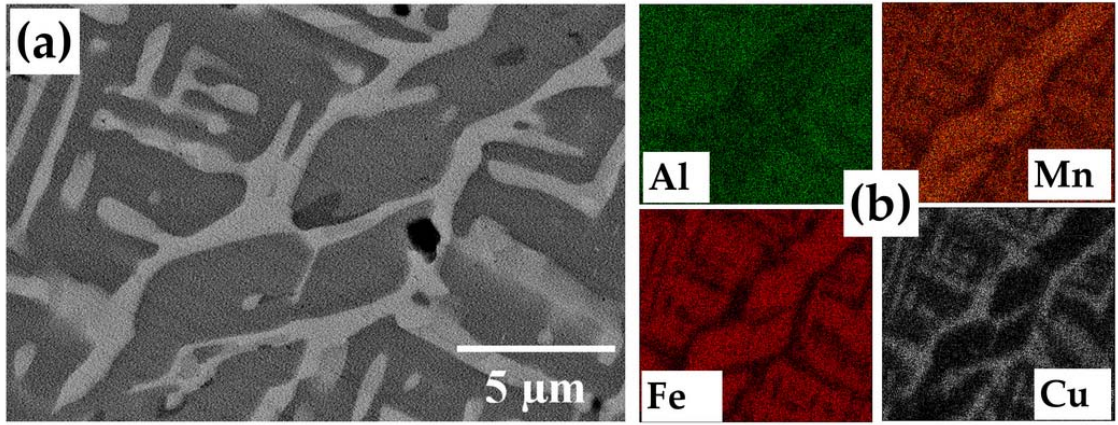

Figure 6: (a) Back-scattered electron (BSE) image and (b) the corresponding elemental distributions obtained from EDS analysis in the AlCuFeMn MPCA.

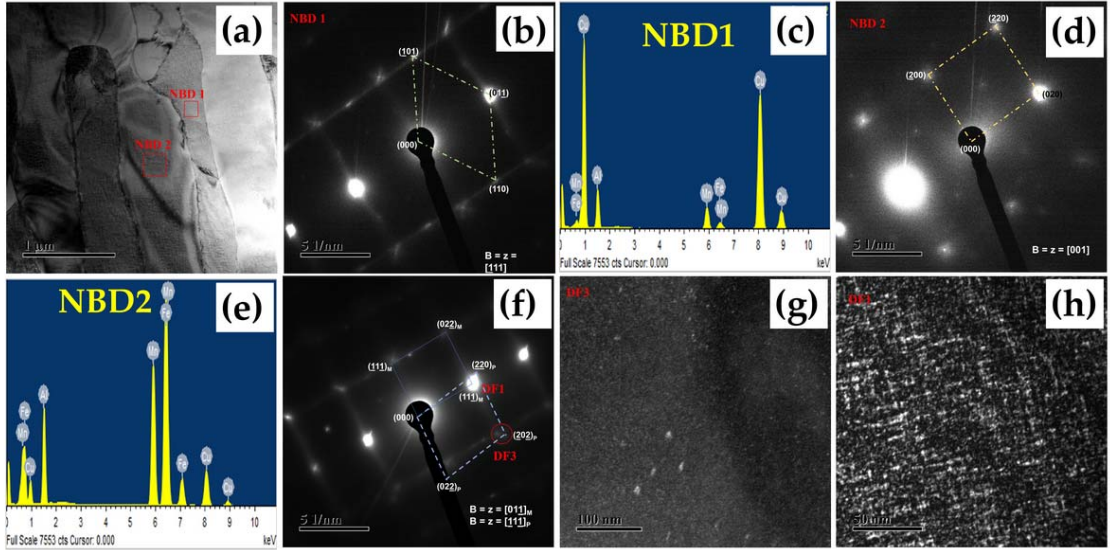

Figure 7: TEM micrograph (a), SADP and corresponding EDS of Cu-rich region (b-c), Fe-Mn-rich region (d-e), SADP pattern of nano-sized second phase $L1_2$ within Fe-Mn-rich region (f), and corresponding dark-field images (g-h) in AlCuFeMn MPCA. 'M' belongs to matrix and 'P' belongs to precipitate.